\documentclass{article}
\usepackage{fullpage}
\usepackage{graphics,psfrag,epsfig}
\usepackage{amsfonts,latexsym,eucal,amsmath,amsthm,amssymb}

\begin{document}

\newcommand{\rum}{\rule{0.5pt}{0pt}}
\newcommand{\rub}{\rule{1pt}{0pt}}
\newcommand{\rim}{\rule{0.3pt}{0pt}}
\newcommand{\numtimes}{\mbox{\raisebox{1.5pt}{${\scriptscriptstyle \times}$}}}
\newcommand{\optprog}[2]
{%
  \noindent\mbox{}\\[0cm]
  \noindent\fbox{%
  \begin{minipage}{0.955\linewidth}
    \mbox{}\\[-0.5cm]
    #1\\[#2]
  \end{minipage}
  }
  \noindent\mbox{}\\[-0.2cm]
}

\renewcommand{\refname}{References}

\twocolumn[%
\begin{center}
{\Large\bf Machine-Checked Formalization of Earlier Arguments\\ on $\mathbb{P}$ versus $\mathbb{NP}$ Using Isabelle/HOL \rule{0pt}{13pt}}\par
\bigskip
Craig Alan Feinstein \\ {\small\it 2712 Willow Glen Drive, Baltimore, Maryland
21209\rule{0pt}{13pt}}\\ \raisebox{-1pt}{\footnotesize E-mail: cafeinst@msn.com,
BS"D}\par
\bigskip\smallskip
{\small\parbox{11cm}{%
\bigskip \noindent \textbf{Abstract:} This letter revisits an earlier argument concerning
$\mathbb{P}$ versus $\mathbb{NP}$ based on the SUBSET-SUM problem
and examines its formalization in Isabelle/HOL.
The formal development clarifies the argument’s logical structure
by separating its deductive combinatorial core
from the broader universality principle required
to extend it to all exact deterministic algorithms.

\bigskip \noindent \textbf{Disclaimer:} This article was authored
by Craig Alan Feinstein in his private capacity. No official support or endorsement by
the U.S. Government is intended or should be inferred.\rule[0pt]{0pt}{0pt}}}
\bigskip
\end{center}]{%

\section*{1. Introduction}

In earlier articles published in \textit{Progress in Physics}
I presented information-theoretic arguments concerning
the Collatz $3n+1$ conjecture and the $\mathbb{P}$ versus $\mathbb{NP}$
problem \cite{CF1,CF2,CF3}. Almost twenty years after my first article,
I undertook a formalization of portions of those arguments in Isabelle/HOL \cite{I26}.

The Collatz development formalized without conceptual difficulty:
definitions were made precise and the reasoning
was expressed in machine-checked deductive form \cite{CF4}.
The SUBSET-SUM lower-bound argument proved more subtle \cite{CF5}.
Its combinatorial component formalizes directly,
but extending that component to a universal lower bound
over all exact deterministic algorithms does not.

The purpose of this letter is to describe what emerged from the
formalization — namely, a separation between the deductive
combinatorial core of the argument and the broader general
principle required to extend it to algorithmic universality.

\section*{2. The SUBSET-SUM argument}

The SUBSET-SUM decision problem asks:
given integers $s_1,\dots,s_n$ and a target $t$,
does there exist a subset $I \subseteq \{1,\dots,n\}$
such that
\[
\sum_{i \,\in \,I} s_i = t \, ?
\]

\noindent There are $2^n$ possible subsets.
Fix $k$ and partition the index set into
$\{1,\dots,k\}$ and $\{k+1,\dots,n\}$.
This induces the families
\[
\begin{array}{l}
\displaystyle
L_k =
\left\{
\sum_{i \,\in\, I^+} s_i :
I^+ \subseteq \{1,\dots,k\}
\right\},\\[16pt]
\displaystyle
R_k =
\left\{
t - \sum_{i \,\in\, I^-} s_i :
I^- \subseteq \{k+1,\dots,n\}
\right\}.
\end{array}
\]

\noindent The verification condition is equivalent to
\[
L_k \cap R_k \neq \varnothing.
\]

\noindent If the subset-sums on each side are distinct, then
\[
|\rule{1pt}{0pt}L_k\rule{1pt}{0pt}| = 2^k,
\qquad
|\rule{1pt}{0pt}R_k\rule{1pt}{0pt}| = 2^{n-k}.
\]

\noindent The quantity $2^k + 2^{n-k}$ is minimized when
$k = \lfloor n/2 \rfloor$, yielding $\Theta\rule{1pt}{0pt}(2^{n/2})$.
All of these identities formalize directly in Isabelle/HOL.
The combinatorial structure is explicit and fully verified.

\section*{3. The missing step}

What does not formalize is the further claim that this
combinatorial structure alone entails exponential
worst-case behavior for all exact deterministic algorithms.
In \cite{CF3}, the exponential conclusion was motivated by
the observation that the verification equation does not appear
to admit a fundamentally simpler equivalent formulation.
All known algebraic rearrangements and equivalent reformulations
preserve the exponential number of candidate values.

The transition from this observation to a universal lower bound,
however, is not a deductive consequence of the combinatorial
identities themselves. It is an additional general principle.
The proof assistant makes this distinction explicit.
It verifies the combinatorial theorems,
while requiring universal claims to be stated
as explicit assumptions or independently proved results.
In particular, a universal conclusion cannot be obtained
solely from the absence of a known simplification.
This does not alter the substance of the argument;
it clarifies its logical structure:
the split analysis is deductive,
whereas the universal lower-bound claim
rests on an additional principle.

It is natural to conjecture that the verification equation,
in its standard form, admits no genuine simplification.
All familiar algebraic rearrangements preserve the exponential
proliferation of candidate values revealed by the split construction.
In this sense, the equation appears structurally rigid.
One might hope to prove such rigidity formally.
A theorem of this kind would show that, within a fixed formal
framework, no nontrivial simplification can be derived.
Yet even such a result would not completely settle the issue.
It would only show that no simplification is provable there,
not that alternative formulations of SUBSET-SUM cannot exist.
The essential question is therefore not merely whether this
verification equation can be simplified, but whether every formulation
of SUBSET-SUM must exhibit the same combinatorial structure.

\section*{4. The LR-read principle}

Within the formal development,
the additional step can be isolated as an explicit assumption.

\paragraph{LR-read principle (informal statement).}
Any exact deterministic algorithm for SUBSET-SUM
must, in the worst case,
distinguish values arising from both sides
of some partition of the index set
into subsets of sizes $k$ and $n-k$.

\bigskip
\noindent Equivalently, no exact deterministic algorithm can decide
\[
L_k \cap R_k \neq \varnothing\,,
\]
while avoiding distinction among
exponentially many induced values.

Under this principle,
the combinatorial structure yields
an exponential lower bound.
Without it, the counting identities alone do not imply
a universal lower bound.
The original argument implicitly relied on this principle.
The formal development makes it explicit.

\section*{5. Deductive core and universality}

The formal development separates the argument
into two logically distinct components.

\paragraph{Deductive core.}
The split construction, cardinality identities, and minimization.
These are exact theorems and fully machine-verified.

\paragraph{Universality step.}
The assertion that every exact deterministic algorithm
must respect the distinction captured by LR-read.

\bigskip \noindent The proof assistant does not alter the substance of the argument;
it makes explicit which parts are formally derived
and which depend on an additional principle.

\section*{Conclusion}

This development does not establish
$\mathbb{P} \neq \mathbb{NP}$
as an unconditional deductive theorem.
Rather, it shows that under the additional modelling
principle introduced here, the combinatorial structure of
SUBSET-SUM yields an exponential lower bound.
The central question becomes:

\bigskip\noindent \textit{Does there exist an exact deterministic algorithm
for SUBSET-SUM whose structure fundament\-al\-ly avoids
the exponential distinction revealed by the split analysis?}

\bigskip\noindent The formalization makes explicit the precise point
at which the argument moves beyond deduction.
Accordingly, the issue reduces to whether the verification equation
admits an equivalent formulation that avoids the exponential
structure revealed by the split analysis.

It is intuitively plausible that no such equivalent formulation exists,
since all known reformulations preserve the same combinatorial structure.
If such a formulation exists, the lower-bound argument would not apply.
If no such formulation exists, then the argument would extend
to all exact deterministic algorithms, and the exponential
distinction would follow.

More broadly, this clarification applies not only to the present
SUBSET-SUM argument but also to the author's earlier work on
complexity-theoretic questions. In each case, the deductive core
is accompanied by additional assumptions that are intuitively
plausible but not themselves derived as formal theorems \cite{CF1,CF2,CF6,CF7}.

}

\end{document}